# Classification of COVID-19 Patients with their Severity Level from Chest CT Scans using Transfer Learning

Mansi Gupta[1], Aman Swaraj[2], Karan Verma[1]

National Institute of Technology Delhi[1], Indian Institute of Technology, Roorkee[2]
202211011@nitdelhi.ac.in[1], aman_s@cs.iitr.ac.in[2], karanverma@nitdelhi.ac.in[1]

## ABSTRACT

**Background and Objective:**

During pandemics, the use of artificial intelligence (AI) approaches combined with biomedical science play a significant role in reducing the burden on the healthcare systems and physicians. The rapid increment in cases of COVID-19 has led to an increase in demand for hospital beds and other medical equipment. However, since medical facilities are limited, it is recommended to diagnose patients as per severity of the infection. Keeping this in mind, we share our research in detecting COVID-19 as well as assessing its severity using chest-CT scans and Deep Learning pretrained models.

**Dataset:**

We have collected a total of 1966 CT Scan images for three different class labels, namely, Non-COVID, Severe COVID, and Non-Severe COVID, out of which 714 CT images belong to Non-COVID category, 713 CT images are of Non-Severe COVID category and 539 CT images are of Severe COVID category.

**Methods:**

All of the images are initially pre-processed using the Contrast Limited Histogram Equalization (CLAHE) approach. The pre-processed images are then fed into the VGG-16 network for extracting features. Finally, the retrieved characteristics are categorized and the accuracy is evaluated using a support vector machine (SVM) with 10-fold cross validation (CV).

**Result and Conclusion:**

In our study we have combined well-known strategies for pre-processing, feature extraction, and classification which bring us to a remarkable success rate of disease and its severity recognition with an accuracy of 96.05% (97.7% for Non-Severe COVID-19 images and 93% for Severe COVID-19 images). Our model can therefore help radiologists detect COVID-19 and the extent of its severity.

# 1. INTRODUCTION

In the year 2019 December 31, a mystery case of pneumonia was identified in Wuhan, China. This unique coronavirus (2019- nCoV) was discovered as the causative culprit, that was later called COVID-19 by world health organization on 7 January, 2020 [1]. The Coronavirus (CoV) is a member of the Coronaviridae family which is a family of virus, which causes symptoms that ranges from the ordinary cold to a severe illness of respiratory system. The severity of the illness can lead to respiratory distress syndrome (RDS) and sometimes even death [2]. Owing to its rapid growth, WHO declared it as global pandemic in March 2020 [3]. Presently, more than 500 million cases of COVID-19 with 5 million deaths have been reported worldwide. Especially, in densely populated countries such as India, it has become a challenge for the state to cope up with the growing number of cases since the second wave of COVID-19 hit the country in April, 2021. Figure 1 shows the rise of covid cases in recent times in India [4].

In our earlier work [5, 6], we presented a forecasting model to predict the number of covid cases. However, forecasting models can give a general idea about the growth of the pandemic, but it can't diagnose specific patients per se. Therefore, to prevent COVID-19 from rampaging the society, it becomes equally crucial to diagnose patients in early stage [7]. Subsequently, medical facilities have come up with certain laboratory identification tools that can detect presence of COVID-19, such as real-time reverse transcription-polymerase chain reaction (RT-PCR) [8] and isothermal nucleic acid amplification technology.

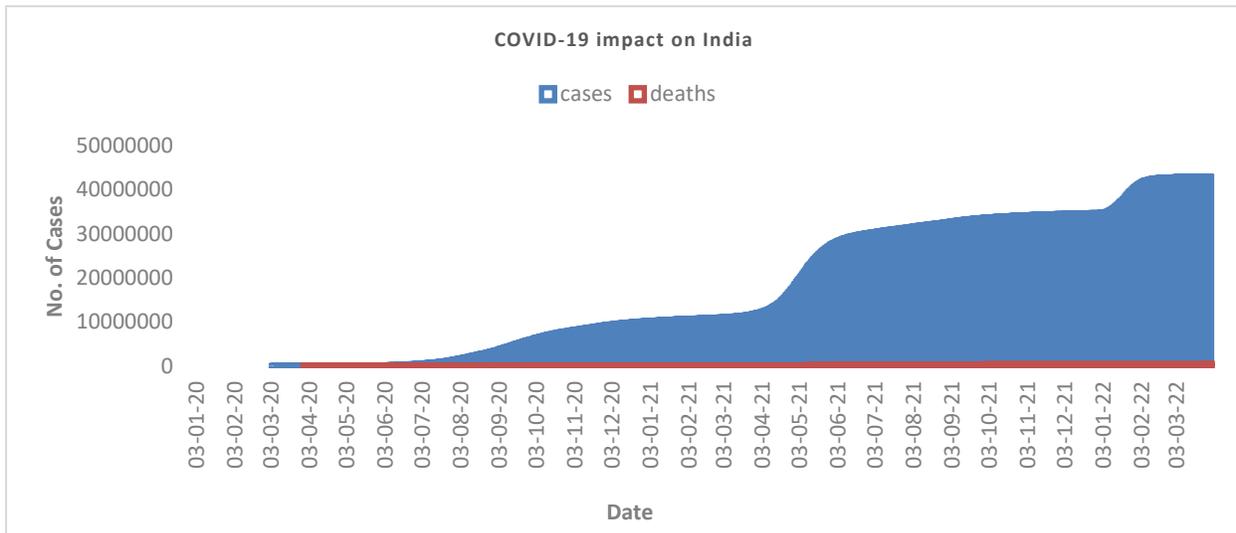

*Fig 1. Total confirmed cases and deaths in India till April 1, 2022*

However, RT-PCR has certain drawbacks, such as insufficient viral material in specimens collected from suspected patients. Also, it is time and resource consuming as in several cases to confirm the test it may be performed two or more times and above that, it can't indicate the severity of covid in patients.

Many studies have compared CT scans to RT-PCR and shown that CT scans are more important for COVID-19 diagnosis [10-12]. According to a recent study, Tao et al. [11] used both of the techniques to compare their respective diagnostic performance as well as reliability for COVID-19 and discovered that out of 1014 patients, 60% RT-PCR findings were positive whereas 88% chest CT scans showed positive results, indicating that latter technique shows a great sensitivity for detection of the COVID-19. The ability of a CT scan to identify the specific symptoms in the chest such as peripheral multifocal ground-glass opacities and consolidations (Fig. 2) has made it an important tool for detection of COVID-19 in early stages along with its severity assessment in many countries [9].

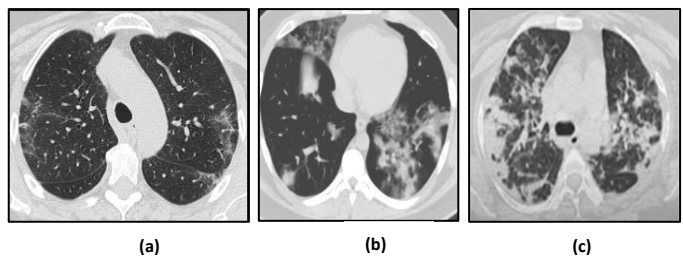

(a)     (b)     (c)

*Fig2. Evolution of COVID-19 Symptoms in Chest (a) Early Stage: Ground Glass Opacity (b) Progressive Stage: Ground Glass Opacity increases in shape and size and changes to Consolidation (c) Peak Stage: Dense Consolidation*

Often during the peak phase of a covid wave, cases see a surge and, in such situations, Manual detection with Severity evaluation on basis of scans becomes an endeavour that requires a significant amount of time and energy, and may cause a delay in the isolation and effective treatment of the patient. Subsequently researchers have adopted Artificial Intelligence (AI) techniques for diagnosing X-rays and CT scans.

In recent times, artificial intelligence (AI) has gained much popularity in advancing biomedical research [13, 14]. Deep learning techniques in particular have worked out very well in case of medical imaging involving a wide range of diseases such as breast cancer, skin cancer, brain tumor etc. [15-18]. Therefore, it would be promising to say that a deep learning-based approach for analyzing chest CT scan images can supplement conventional tools in detecting COVID-19 symptoms along with its severity assessment. With this motivation we present our work on COVID-19 severity assessment on chest CT scan images using deep learning models. Rest of the paper is organized as follows- In section 2, we describe earlier work done in this domain. Section 3 talks about methodology and section 4 depicts results of our proposed methodology. Finally, in section 5 we conclude with potential future work.

## 2. Related Work

This section talks at length about various methodologies adopted by researchers in this domain. First in 2.1, we talk about the detection task and then in section 2.2 we describe work done in the severity assessment area.

**2.1. Detection of COVID-19 using chest CT Images**

Since the inception of COVID in December 2019, several studies have been proposed for detecting the virus and severity of the infection via CT scan images and AI techniques. However, due to data scarcity, most researchers have used transfer learning techniques as it provides significant results with limited-sized

Dataset and also reduces the time required for training the model. In our previous work, we also classified COVID-19 X-ray images based on transfer learning model VGG16 [19].

In this connection, Pathak Y. et al. [20] presented, an approach based on transfer learning to identify COVID-19 using CT images with ResNet50 as the transfer learning model to extract the features and a 2D CNN based model for classification. The suggested system has acquired accuracy of 93.01% after 10-fold-CV evaluation on 413 COVID samples and 439 Non COVID samples. Vruddhi et al. [21] designed a CTnet-10 model with an accuracy of 82.1% and also used different transfer learning techniques like DenseNet169, VGG-16, VGG19, ResNet-50 and InceptionV3 for COVID-19 detection and concluded that VGG19 outperformed other models with accuracy of 94.5%. S. V. Kogilavani et al. [22] also utilized different transfer learning methods like VGG16, DenseNet121, MobileNet, NASNet, Xception, and EfficientNet in their proposed work and observed that VGG16 provided the best accuracy of 97.68% as compared to other architectures.

Similarly, Jaiswal et al [23] designed a DTL model for the classification of COVID by testing two pre-trained models DenseNet 201 and VGG16 on a testing dataset of SARS-Cov-2 and achieved accuracy of 96% and 95.45% respectively thus they selected DenseNet as best model for their work.

Arpita Halder et al. [24] proposed a 2-D Deep Learning framework known as KarNet which used DenseNet201, MobileNet, ResNet50 and VGG16 as its backbone and combined with additional layers to test the performance of each model independently on both original (unaugmented) and altered (augmented) datasets, and found that of the 4 pre-trained models used in KarNet, the one that is incorporated DenseNet201 beat the others with a 97% accuracy. Some of the prominent works in this direction are presented in table 1.

*Table 1: some of the prominent works in covid-19 detection using CT scan images*

| Reference | Data structure and size | Best model structure(s) | Accuracy |
|---|---|---|---|
| *Pathak et al. [20]* | 413 COVID-19<br>439 non-COVID | ResNet50 as feature extractor with CNN as classifier. | 93.01% |
| *Vruddhi et al. [21]* | 349 COVID-19<br>493 non-COVID | VGG-19 | 94.5% |
| *Jaiswal et al. [22]* | 1,262 COVID-19<br>1,230 non-COVID-19 | DenseNet201 | 96.25% |
| *S. V. Kogilavani et al [23]* | 1252 COVID-19<br>1229 non-COVID | VGG-16 | 97.68% |
| *Arpita Halder et al. [24]* | 1252 COVID-19<br>1229 non-COVID | KarNet (DenseNet201, MobileNet, ResNet50 and VGG16 as backbone with DenseNet as best pre-trained model) | 97% |
| *Ying Song et. al. [25]* | 777 COVID-19<br>505 Pneumonia<br>708 Normal | DRENet (ResNet50 and FPN for feature extraction and multilayer perceptron for classification) | 86% |

| Reference | Data structure and size | Best model structure(s) | Performance |
|---|---|---|---|
| *Varan et. al. [26]* | 853 COVID-19 253 Normal | ReCOV-101 (deep CNN model with ResNet-101 as backbone) with segmentation | 94.9% |
| *Parisa gifani et. al. [27]* | 349 COVID-19 397 Non COVID-19 | Ensemble method relies on majority voting of the combination of EfficientNetB3, Inception_resnet_v2, EfficientNetB0, EfficientNetB5 and Xception pre-trained models outputs | 85% |
| *Aswathy A.L et al. [28]* | 760 COVID-19 by augmentation 736 non-COVID-19 | ResNet-50 model | 98.5% |

**2.2. Assessment of COVID-19 Patients on basis of disease Severity using CT Images**

Although majority of the work has been done with respect to straight forward COVID-19 classification, severity-based classification has also caught researcher's attention significantly. In this regard, Zhenyu Tang et al. [25] used Random Forest (RF) machine learning model for severity evaluation (severe or non-severe) of COVID patients by feeding the model with 63 quantitative features that comprises the total lung infection volume/ratio as well as ground-glass opacity (GGO) areas and they have also examined at the severity-related parameters of the resultant assessment model in their research with 87.5% accuracy.

Feng et al. [30] proposed a Lesion Encoder system consisting of U-Net module and Recurrent Neural Networks (RNN) that automatically detected lesions in a CT scan, assessed severity, as well as predicted the disease progression with an accuracy of 94%.

L. Xiao et al. [31] proposed a Deep Learning based model using Multiple Instance Learning and a Residual Convolutional Neural Network (ResNet34) on a dataset of 408 confirmed COVID-19 patients and achieved 81.9% testing accuracy. Zekuan Yu et al. [32] used transfer learning technique DenseNet201for feature extraction and SVM classifier for Severity assessment of Covid-19 and achieved accuracy of 95.34%. Some of the major work in severity detection is mentioned in table 2.

*Table 2: A Comparitive study of different techniques provided to predict severity of Covid-19 patients.*

| Reference | Data structure and size | Best model structure(s) | Performance |
|---|---|---|---|
| *Aswathy A.L et al. [28]* | 349 CT images of 216 COVID-19 patients (Augmented) | Densenet-201 and ResNet-50 model for extracting features and back propagation technique for classification | 97.84% |
| *Zhenyu Tang et al. [29]* | 55 Severe 121 non-Severe (CT images) | Random Forest (RF) Model | 87.5% |
| *Feng et al. [30]* | 22 Severe 320 non-Severe (CT images) | Lesion Encoder System (U-Net module and RNN) | 94% |
| *L. Xiao et al.* [31] | 93 Severe 305 non-Severe | Multiple Instance Learning and a Residual Convolutional Neural Network (ResNet34) | 81.9% |
| *Zekuan Yu et al.* [32] | 246 Severe 483 non-Severe (CT images) | DenseNet-201 with cubic SVM model | 95.34% |
| *Kelei He et al.* [33] | 666 COVID-19 CT Images (51 severe patients 191 non-severe patients) | M2UNet (multi-task multi-instance UNet) | 98.5% |
| *Nasrin et. al.* [34] | 42 severe cases 120 moderate, 563 mild, 231 non covid | GLRLMS as feature extractor and RF as classifier | 90.95% |
| *Zhang et. al.* [35] | 661 CT images of 24 COVID-19 patients | UNet | 91.6% |

It is quite evident from both the tables that work in the direction of severity detection has scope for improvement compared to task of mere covid detection which has achieved significant performance. Further, in table 2, it is quite evident that most of the dataset used are imbalanced. We overcome this issue in our work. Also, our proposed model will not just detect the covid infected patient but also classify them into severe and non-severe cases.

## 3. METHODOLOGY

The diagram in Fig. 5 depicts a high-level overview of our suggested system. It consists of four stages: collection of dataset, pre-processing of images, extraction of features, and classification.

### 3.1. Dataset Description

For this research, authors have used the SARS CoV2 chest-CT scan dataset [36]; available publicly, contains 1252 images of patients infected with COVID-19 and 1229 scans of individuals who were not infected by COVID-19 (Table 3). Further for severity classification, we have divided 1252 CT images of COVID-19 positive into 539 severe and 713 non-severe CT images with the help of well experienced radiologist and used 714 non-covid CT images out of 1230 non-covid CT images for achieving a balanced class distribution (Table 4). Figure 3 broadly displays the difference between the three categories.

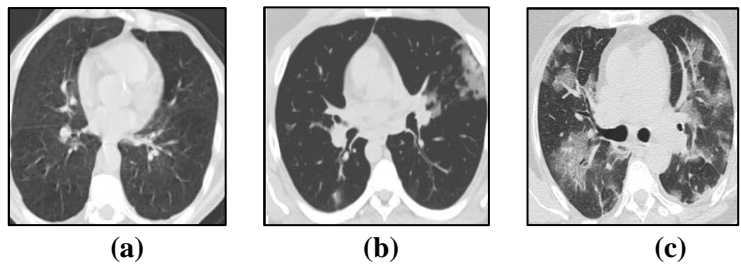

(a)      (b)      (c)

*Fig. 3: Some sample CT Scan images of normal and COVID-19 patients (a) Normal Patient (b) COVID -19 patient (Non-Severe) (c) COVID -19 patient (Severe)*

*Table 4: Class distribution of Sampled chest-CT scan dataset used for our*

| Class | Images |
|---|---|
| *Non-COVID* | 714 |
| *Non-Severe COVID* | 713 |
| *Severe COVID* | 539 |
| *Total* | 1966 |

## 3.2. Data Pre-processing

Before feature extraction, all the images were resized to 224x224 dimension and denoised via a median filter. Further, to improve the quality and information content of image, we applied Contrast Limited Adaptive Histogram Equalization (CLAHE) method.

*3.2.1. Image Resizing*

In computer vision Resizing of images is an important preprocessing step as a large number of transfer learning model designs need that input size of the images is uniform but our unprocessed images varies in size. Pre-trained model receives inputs of a particular size, in this case being 224x224. So, all the images were resized appropriately.

*3.2.2. Image Denoising*

When it comes to processing of images, image denoising is a major challenge. Image noise can be created by a variety of inherent or extrinsic situations that are challenging to cope with. So, to remove the noise present in the collected images we have used median filter. Median filtering is a nonlinear method for minimising impulsive noise, sometimes known as salt-and-pepper noise.

*3.2.3. Image Enhancement*

Image enhancement helps to improve the quality and information content of original data thus helps in extracting the features efficiently in computer vision. For improving the visibility of our image, we have used CLAHE. Fig 4 shows the difference in images before and after applying the image enhancement technique.

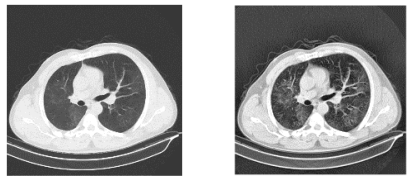

*Fig 4: From left, original CT scan and after applying CLAHE*

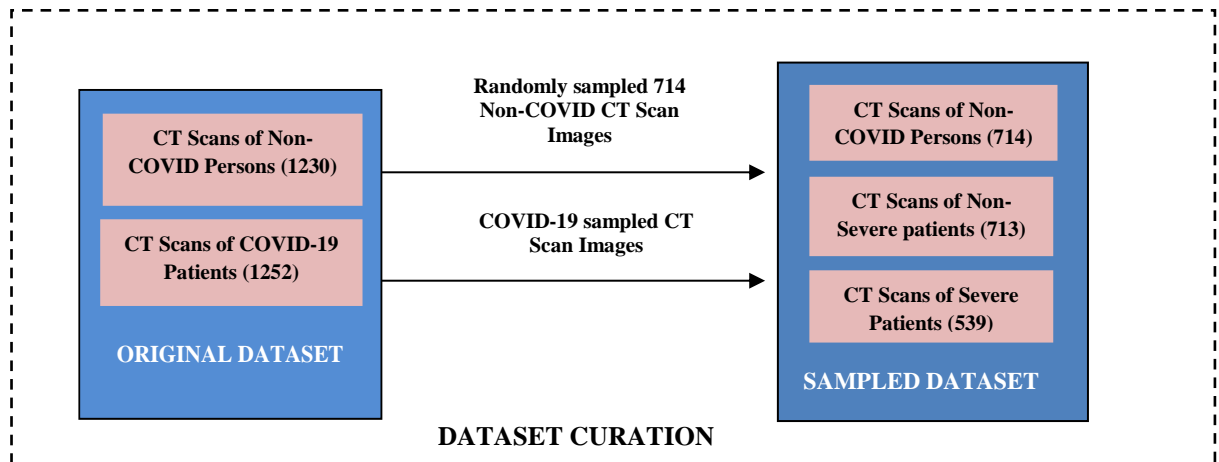

**DATASET CURATION**

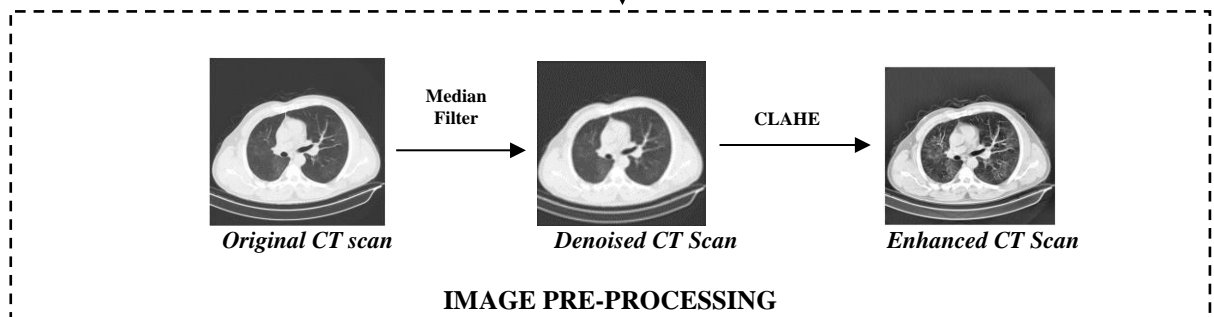

**IMAGE PRE-PROCESSING**

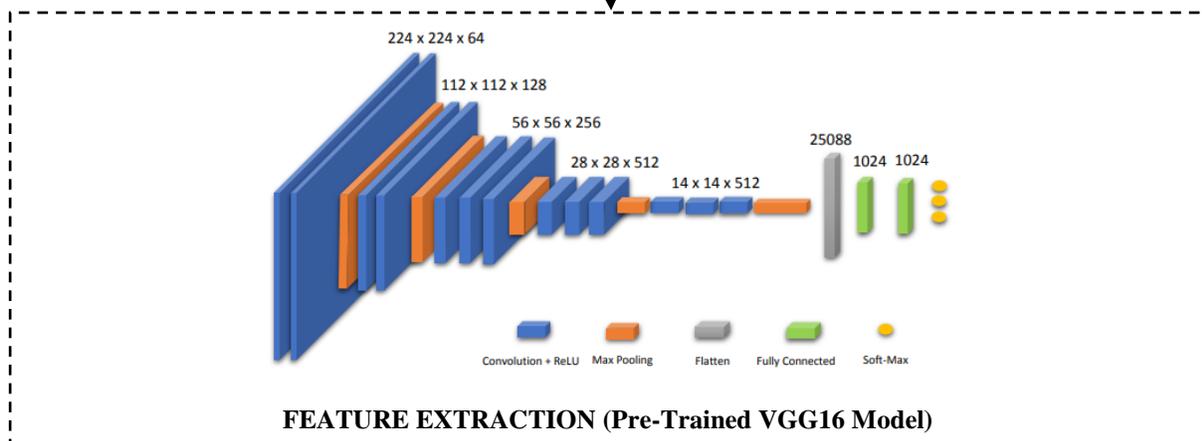

**FEATURE EXTRACTION (Pre-Trained VGG16 Model)**

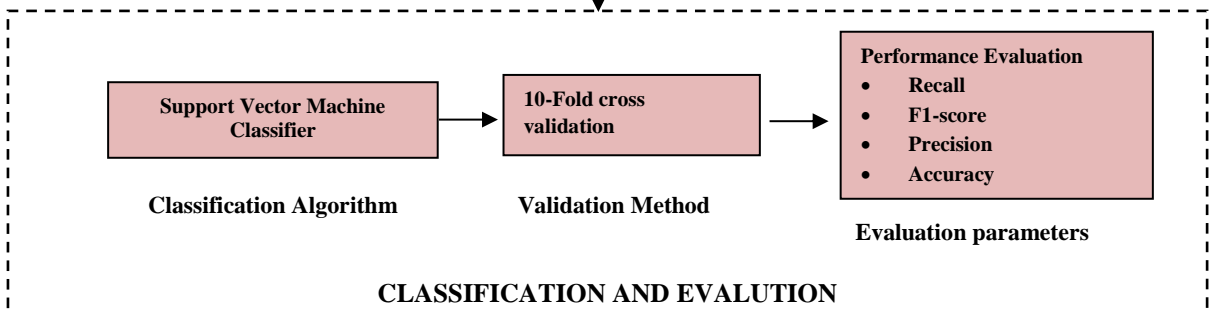

**CLASSIFICATION AND EVALUTION**

*Fig 5: Block Diagram of Proposed framework*

### 3.3. Feature Extraction

Deep learning is well-known for its ability to efficiently represent image features. Thus, it has been successfully applied to a variety of image analysis and classification jobs. But for applying Deep Learning CNN, the dataset required must be in extensive amount and also it takes long time to train the deep CNN model. So, to overcome the mentioned problem we have applied transfer learning.

Transfer learning is a deep learning approach in which a neural network model is trained on an issue that is similar to the one that has to be solved. It includes Models that have been pre-trained for common benchmark datasets for computer vision, such as the ImageNet image recognition tasks. The best models can then be utilised directly or combined into a new model to solve our own computer vision challenges.

So, for our study we have used different pre-trained models such as Resnet50 [38], VGG19, VGG16 [39], DenseNet121 [40] for extracting the features from the CT images present in our sampled dataset. These CNN-based models that have been pre-trained were chosen according to their stated accuracy. We have achieved the best results by using VGG16 pre-trained model (Fig 6, Table 5).

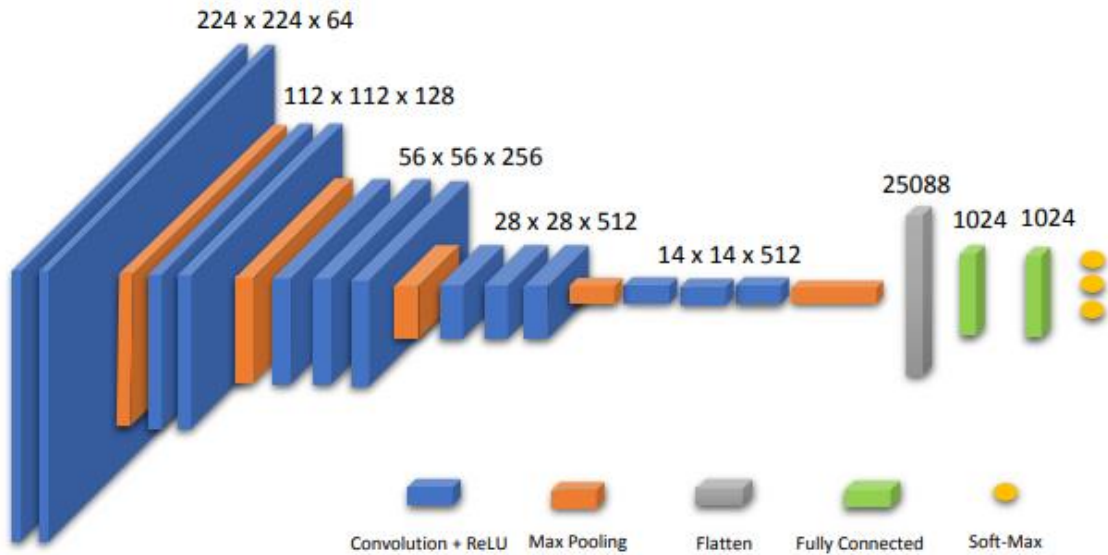

*Fig 6: VGG-16 network architecture for feature extraction*

*Table 5: The network structure of VGG16 transfer learning pre-trained model used in this paper.*

| Layer | Kernel Size | Stride | Padding | Kernel | Activation |
|---|---|---|---|---|---|

| | | | | Initialization | |
|---|---|---|---|---|---|
| Conv2D | 3x3 | 1x1 | Same | he_normal | Relu |
| Conv2D | 3x3 | 1x1 | Same | he_normal | Relu |
| Maxpool | 2x2 | 2x2 | Same | None | None |
| Conv2D | 3x3 | 1x1 | Same | he_normal | Relu |
| Conv2D | 3x3 | 1x1 | Same | he_normal | Relu |
| Maxpool | 2x2 | 2x2 | Same | None | None |
| Conv2D | 3x3 | 1x1 | Same | he_normal | Relu |
| Conv2D | 3x3 | 1x1 | Same | he_normal | Relu |
| Conv2D | 3x3 | 1x1 | Same | he_normal | Relu |
| Maxpool | 2x2 | 2x2 | Same | None | None |
| Conv2D | 3x3 | 1x1 | Same | he_normal | Relu |
| Conv2D | 3x3 | 1x1 | Same | he_normal | Relu |
| Conv2D | 3x3 | 1x1 | Same | he_normal | Relu |
| Maxpool | 3x3 | 1x1 | Same | None | None |
| Conv2D | 3x3 | 1x1 | Same | he_normal | Relu |
| Conv2D | 3x3 | 1x1 | Same | he_normal | Relu |
| Conv2D | 3x3 | 1x1 | Same | he_normal | Relu |
| Maxpool | 3x3 | 1x1 | Same | None | None |

### 3.4. Classification

The majority of Covid-19 detection experiments have employed simply Deep Learning algorithms or transfer learning techniques with fine-tuned fully-linked layer. However, training deep neural network from initial stage or fine-tuning parameters of bespoke fully-linked layers on the top of model that has already been trained necessitates a large quantity of data as well as high-performance computational resources, which are typically unavailable.

As a result, we started by extracting features by simply supplying the "include top" parameter as "False" when loading a model that has been pretrained without the classifier, thus freezing the convolutional base and using it as feature extractor only and then we have added our own classifier, on top of the pretrained model for the classification task.

We applied three distinct machine learning-based categorization methods for our research: the standard SVM, RF and XGBoost (Gradient-Boosting Machine). We have achieved best results with SVM classifier so we have finalized SVM classifier for the classification purpose.

### 3.5. EVALUATION METRICS

To assess the performance of suggested model, authors have used different performance indicators such as: sensitivity (Se) / recall (Re), specificity (Sp), accuracy (ACU), precision (Pr) and F1-score (f1). True-Positive (TP), False-Positive (FP), True-Negative (TN), and False-Negative (FN) parameters were derived from each class of confusion matrix to compute these metrics.

$$S_e = \frac{TP}{TP+FN} \tag{1}$$

$$S_P = \frac{TP}{TN+FP} \tag{2}$$

$$ACC = \frac{TP+TN}{TP+FN+FP+TN} \tag{3}$$

$$Pre = \frac{TP}{TP+FP} \tag{4}$$

$$f1\text{-score} = \frac{2*TP}{2*TP+FN+FP} \tag{5}$$

## 4. Results:

The outcomes of the suggested technique are shown in this section. Confusion matrix obtained for our proposed model is illustrated in Fig 7. From the confusion matrix we can observe that the system can reliably classify the severe and non-severe COVID with an accuracy of 93% and 97.7% respectively. For severe cases, the model has correctly predicted 107 CT images and incorrectly classified 8 CT images as Non-Severe. Similarly, for non-severe cases, our model has correctly predicted 127 CT images and incorrectly classified 3 CT images as Severe Covid.

We have also applied 10-Fold-CV for our model to evaluate each fold's performance metrics and its result is illustrated in Table 6. The average accuracy obtained for the proposed model after 10-fold cross validation is 96.05% with f1 score 95.96%. Table 7 shows the various performance metrics calculated for each class in multiclass classification. We've additionally employed the area under ROC curve as a performance metric.

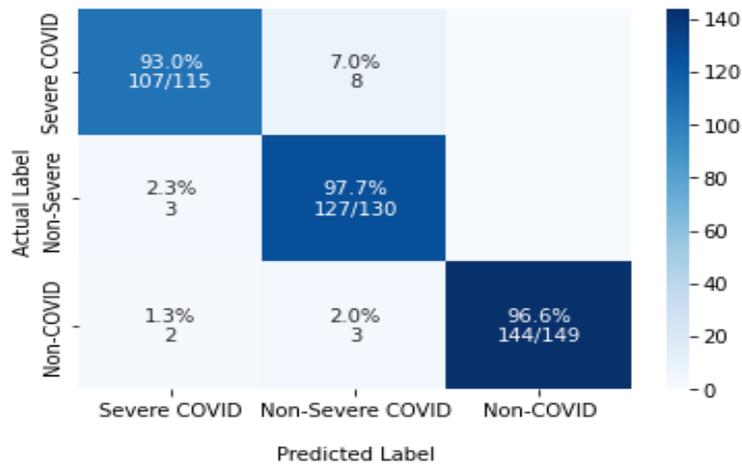

*Fig 7: Confusion matrix for Non-Covid and Covid severity classification of the CT volumes*

Table 6: *Support, accuracy, precision, recall and F1-score values of the proposed model for each fold*

| Fold | Support | Accuracy | Precision | Recall | F1-Score |
|---|---|---|---|---|---|
| 1 | 394 | 95.56 | 95.67 | 95.50 | 95.56 |
| 2 | 394 | 96.20 | 95.95 | 95.91 | 96.20 |
| 3 | 394 | 96.81 | 96.72 | 96.82 | 96.79 |
| 4 | 394 | 95.54 | 95.37 | 95.08 | 95.54 |
| 5 | 394 | 94.26 | 95.13 | 93.28 | 94.17 |
| 6 | 394 | 94.90 | 95.24 | 94.71 | 94.15 |
| 7 | 394 | 96.17 | 95.95 | 96.32 | 96.17 |
| 8 | 394 | 97.45 | 97.47 | 97.47 | 97.45 |
| 9 | 394 | 96.17 | 96.52 | 95.70 | 96.17 |
| 10 | 394 | 97.45 | 97.38 | 97.63 | 97.44 |
| **Average** | **394** | **96.05** | **96.14** | **95.84** | **95.96** |

Table 7: *Support, accuracy, precision, recall and F1-score values of the proposed model for each class in multiclass classification*

| Class | Support | Accuracy | Precision | Recall | F1-Score |
|---|---|---|---|---|---|
| *Non-COVID* | 149 | 96.6 | 100.0 | 96.64 | 98.29 |
| *Severe COVID* | 115 | 93.0 | 95.53 | 93.04 | 94.27 |
| *Non-Severe COVID* | 130 | 97.7 | 92.02 | 97.69 | 94.77 |
| *Average* | **394** | **95.76** | **95.85** | **95.79** | **95.78** |

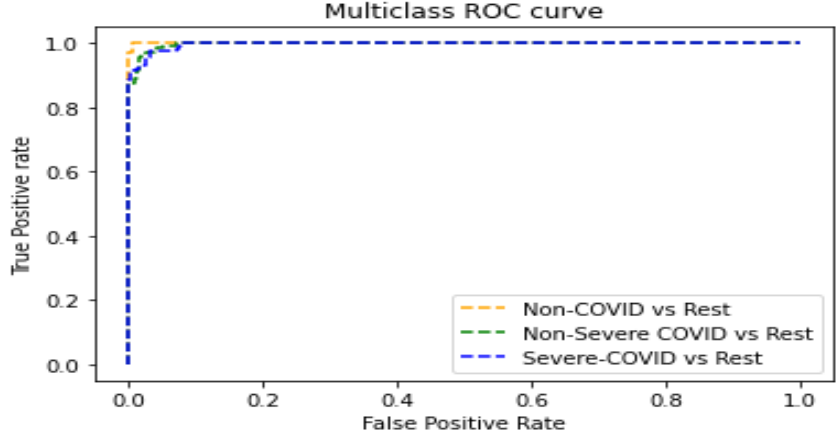

*Fig 8: ROC curve showing area under each class for multiclass classification*

## 5. Discussion and Conclusion

Due to the rapid growth of coronavirus around the world causing millions of infections and deaths it become necessary to diagnose the disease quickly and reliably according to its severity level. The diagnosis techniques like RT-PCR that are used currently is not so efficient as it takes long time and has less sensitivity. Recent research showed that Chest radiographs like chest-CT scan and Chest X-Ray can be used to properly cure COVID-19. We attempted to uncover main problems and limitations in previous studies related to COVID-19 and its severity identification and categorization utilising chest CT scan images during the course of this research. We will analyse the limitations of the existing research described in section 2 in this part.

For this study, we sampled a total of 1966 CT images (714 Non-COVID-19, 713 Non-Severe COVID-19, and 539 Severe COVID-19) from the SARS-CoV-2 CT scan dataset. Most of the studies that we have discussed in section 2 have used subset of this dataset or the sources of this dataset for their work. Table 8 highlight a comparative analysis of our work with some of the recent work in this direction.

*Table 8: Comparison of proposed model with state-of-the-art severity detection methods*

| Reference | Data structure and size | Performance |
|---|---|---|
| *Aswathy A.L et al. [28]* | 349 CT images of 216 COVID-19 patients (Augmented) | 97.84% |
| *Zhenyu Tang et al. [29]* | 55 Severe<br>121 non-Severe (CT images) | 87.5% |
| *Feng et al. [30]* | 22 Severe<br>320 non-Severe (CT images) | 94% |
| *L. Xiao et al. [31]* | 93 Severe<br>305 non-Severe | 81.9% |
| *Zekuan Yu et al. [32]* | 246 Severe<br>483 non-Severe (CT images) | 95.34% |
| *Kelei He et al. [33]* | 666 COVID-19 CT Images (51 severe patients 191 non-severe patients) | 98.5% |
| *Nasrin et. al. [34]* | 42 severe cases 120 moderate, 563 mild, 231 non covid | 90.95% |
| *Zhang et. al. [35]* | 661 CT images of 24 COVID-19 patients | 91.6% |
| ***Proposed work*** | **714 non covid; 713 non-severe and 539 severe cases** | **96.05%** |

Besides table 8, some of the other drawbacks of earlier studies mentioned in section 2 with respect to our work include-

- Deep Learning or transfer learning with a fine-tuned fully linked layer may result in data scarcity and increased time consumption.
- All of the existing research mentioned are proposed either to detect Covid or to detect Covid severity but not both at the same time by using single model.
- Also, there is comparatively lack of training data in the studies related to severity prediction (discussed in section 2).
- Studies regarding severity detection have a major drawback of imbalanced dataset.
- They have used augmentation techniques which needs to be validated or it could lead to overfitting and loss of generalization.

In the proposed framework, we attempted to solve all of these issues in the suggested framework. The main advantages of the proposed model are -

- We have used a combination of transfer learning pre-trained model (VGG16) and machine learning model (SVM), by using this we tried to avoid the problem of data scarcity and try to avoid the use of data augmentation techniques (for increasing the dataset).
- Our proposed work helps to detect COVID-19 as well as predict its severity at the same time.
- We have used comparatively more data for our study.
- We have balanced dataset in our study.

The proposed framework obtained an accuracy of 96.6% for the Non-COVID19 class, 97.7% for the Non-Severe COVID19 class, and 93% for the Severe COVID19 class. The average accuracy of our proposed method is 96.05% with f1-Score 95.96% and ROC score 99.78. In comparison to prior research, our study was able to obtain better accuracy with significantly more data. We will now discuss some of our work's limitations and potential opportunities -

- Despite the fact that the present study employed a larger dataset, the number of Covid-19 CT images used for severity identification is still low.
- By segmenting CT scans depending on infected zones, the suggested model's reliability and efficiency may be enhanced.
- The proposed model must be tested on a bigger dataset.